\documentclass[aps,eqsecnum,nofootinbib,superscriptaddress,showpacs]{revtex4}
\usepackage{amsmath}
\usepackage{amsfonts}
\usepackage{amssymb}
\usepackage[dvips]{graphicx}  
\usepackage{amsmath,amsfonts}
\usepackage{color}  
\def\be{\begin{eqnarray}}
\def\ee{\end{eqnarray}}
\def\beq{\begin{equation}}
\def\eeq{\end{equation}}

\def\({\left (}
\def\){\right )}
\def\S{{\cal S}}

\newtheorem{theorem}{Theorem}[section]
\newtheorem{lemma}[theorem]{Lemma}
\newtheorem{proposition}[theorem]{Proposition}
\newtheorem{corollary}[theorem]{Corollary}
\newtheorem{definition}{Definition}[section]
\newtheorem{remark}[theorem]{Remark}

\newcommand{\qed}{\nobreak \ifvmode \relax \else
      \ifdim\lastskip<1.5em \hskip-\lastskip
      \hskip1.5em plus0em minus0.5em \fi \nobreak
      \vrule height0.75em width0.5em depth0.25em\fi}
\newcommand{\qcd}{\begin{flushright} $\Box$ \end{flushright}}
\begin{document}

\title{
Some remarks on marginally trapped surfaces and geodesic incompleteness 
}

\author{I.P. Costa e Silva}
\email{ivanpcs@mtm.ufsc.br}
\affiliation{Department of Mathematics,\\ 
Universidade Federal de Santa Catarina \\88.040-900 Florian\'{o}polis-SC, Brasil}

\date{\today}

\begin{abstract}
In a recent paper \cite{EGP}, Eichmair, Galloway and Pollack have proved a Gannon-Lee-type singularity theorem based on the existence of marginally outer trapped surfaces (MOTS) on noncompact initial data sets for globally hyperbolic spacetimes. However, one might wonder whether the corresponding incomplete geodesics could still be complete in a possible non-globally hyperbolic extension of spacetime. In this note, some variants of that result are given with weaker causality assumptions, thus suggesting that the answer is generically negative, at least if the putative extension has no closed timelike curves. However, unlike in the case of MOTS, on which only the outgoing family of normal geodesics is constrained, we have found it necessary in our proofs to impose also a weak convergence condition on the {\em ingoing} family of normal geodesics. In other words, we consider {\em marginally trapped surfaces} (MTS) in chronological spacetimes, introducing the natural notion of a {\em generic} MTS. In particular, a Hawking-Penrose-type singularity theorem is proven in chronological spacetimes with dimension $n \geq 3$ containing a generic MTS. Such surfaces naturally arise as cross-sections of  quasi-local generalizations of black hole horizons, such as dynamical and trapping horizons. We end with some comments on the existence of MTS in initial data sets. 
              
\end{abstract}
\maketitle

\section{Introduction}\label{sec:intro}

The Hawking-Penrose singularity theorems \cite{BE,HE,oneill,P,penrose} are milestones of Mathematical Relativity. Apart from their fundamental role in shaping our current understanding of (classical aspects of) black holes and the Big Bang, these theorems have motivated the introduction of a number of fundamental techniques in Lorentzian Geometry. A key idea in these theorems is the use of {\em trapped sets} as mathematical models for regions of spacetime undergoing gravitational collapse. The hypotheses in the various singularity theorems then give sufficient, physically motivated conditions for the occurrence of a complete gravitational collapse, geometrically described as nonspacelike geodesic incompleteness of the spacetime manifold.

A seemingly unrelated issue is whether and how does the causal structure, which underpins General Relativity and other geometric theories of gravity, constrain the underlying topology of spacetime. This relationship between the causal and topological structures of spacetimes is clearly of perennial interest, both for Physics and for Geometry.  The now classic singularity theorems proved independently by Gannon \cite{gannon1,gannon2} and Lee \cite{lee} in 1975/6, were among the earliest fruitful attempts to address this question in a mathematically precise, model-independent fashion. As an example of how these theorems relate to previous singularity theorems, one of Gannon's theorems \cite{gannon1} establishes the null geodesic incompleteness of a spacetime $(M,g)$ which satisfies the {\em null convergence condition} [i.e., $\mbox{Ric} (v,v) \geq 0$ for any null vector $v \in TM$, where $\mbox{Ric}$ denotes the Ricci tensor of $(M,g)$], and which admits a smooth, spacelike Cauchy hypersurface $\S \subset M$ which is regular near infinity [a sort of asymptotic flatness requirement] and nonsimply connected. In what follows we shall refer to this result simply as the {\em Gannon-Lee theorem}. We note that Gannon also tried \cite{gannon1} to prove a version of his theorem for {\em chronological spacetimes}, in the spirit of the Hawking-Penrose main theorem in Ref. \cite{penrose}. Unfortunately, as first pointed out in \cite{galloway2}, his proof for this result is flawed, being based on a false assumption (cf. \cite{gannon1}, pg. 2366, second paragraph). (However, a generalized version of the Gannon-Lee theorem does exist for the weaker assumption that spacetime is {\em causally simple} \cite{me}.)    

The intuitive physical picture that emerges from these theorems is that certain nonsimply connected cross-sections of spacetime naturally collapse into a singularity under reasonable physical conditions, at least when quantum effects can be neglected. This scenario is reinforced by a geometric analysis of exact solution of the Einstein field equations such as the Schwarzschild-Kruskal spacetime, and more generally from the study of specific spacetimes containing wormholes. If some weak version of the Cosmic Censorship conjecture holds, then one expects such singularities to occur behind event horizons. Such an expectation is reinforced by the {\em Topological Censorship theorems} \cite{FSW,galloway1,galloway3}, which under conditions similar to those in the Gannon-Lee theorem imply that these nonsimply connected sections and the ensuing singularity must be hidden from active scrutiny (in a suitable sense) by a ``distant observer''. 

One disadvantage of the latter set of theorems lies in the fact that they are {\em global} in nature, in the sense that checking their assumptions concretely requires foreknowledge of the whole spacetime. This is of course a limitation in trying to detect the development of singularities in numerical Relativity, for example. In a recent paper \cite{EGP}, Eichmair, Galloway and Pollack have addressed this issue, treating  Topological Censorship from a initial data perspective. A key point of their analysis is investigating the existence of {\em marginally outer trapped surfaces} (MOTS) in initial data sets, as they find that MOTS naturally form in asymptotically flat initial data sets with non-trivial topology. In addition, they prove that if the existence of such a MOTS $\Sigma$ contained in a noncompact Cauchy hypersurface in spacetime is granted, and this MOTS is moreover ``generic'' in the sense that future and past inextendible null geodesics normal to $\Sigma$ have non-zero tidal acceleration somewhere along them, then null geodesic incompleteness follows. (The null convergence condition is also assumed.) This result can therefore be understood as a Gannon-Lee-type singularity theorem.  

Now, since the latter result primarily addresses initial data sets, attention is restricted in \cite{EGP} to their globally hyperbolic Cauchy developments. (See also \cite{AMMS}, where a singularity theorem in globally hyperbolic spacetimes containing MOTS is proven, with alternative generic conditions on the MOTS.) This is usually enough for numerical Relativity. However, from a more general perspective one naturally wishes to know whether the geodesic incompleteness thus deduced would survive when further, non-globally hyperbolic extensions of spacetime are made. We emphasize that this is far from being a mere appendix to the globally hyperbolic setting, as there may well exist physically relevant, non-globally hyperbolic extensions of a maximal {\em globally hyperbolic} development of a given initial data which is nonspacelike geodesically complete. Domains of dependence of partial Cauchy hypersurfaces in anti-de Sitter spacetime offer a simple illustration of this point. The specific proof given in \cite{EGP} cannot be generalized to this setting without extra input.    

It is the purpose of this note to investigate such singularity theorems with weakened causality condition, namely the non-existence of closed timelike curves. However, we have found it necessary to work with {\em marginally trapped surfaces} (MTS), instead of MOTS. In other words, we impose a sign condition on the ``ingoing'' family of normal geodesics to a closed, codimension two surface, together with a generic condition on this family as well. The study of such surfaces in Mathematical Relativity has been intense in the last ten years or so, in connection with the various quasi-local notions of black hole horizons which have arisen (see, e.g., \cite{AK} for a review). In particular, we prove a Hawking-Penrose type singularity theorem (see \cite{penrose}, and Theorem 2, p. 266 of \cite{HE} - henceforth we refer to this result as {\em the} Hawking-Penrose singularity theorem), in which one assumes, besides the chronological condition, the timelike convergence condition (which is just the strong energy condition for spacetimes satisfying the Einstein field equations), together with the (standard) generic condition on nonspacelike geodesics. Recall that the latter condition means that there are non-zero tidal accelerations along {\em any} inextendible nonspacelike geodesic. (Although this might sound too restrictive, it is well known that this condition can fail only under special circumstances (see, e.g. Chs. 2, 12 and 14 of \cite{BE} for a thorough discussion of the generic condition), so it is widely believed that spacetimes in which it does {\em not} hold hold form, very loosely speaking, ``a zero-measure set'' in the class of physically relevant spacetimes.) Clearly, some sort of genericity condition is needed to infer incompleteness. For instance, by performing simple isometric identifications in Minkowski spacetime one can check that there are even globally hyperbolic spacetimes which are geodesically complete while having compact marginally (outer) trapped surfaces. 

MOTS have been extensively studied in the Mathematical Physics literature, and some existence results for them have been proven \cite{AEM,AMMS,AM,eichmair1,eichmair2}. Apart from their natural appearance in Relativity, they are particularly interesting to geometers because they have a number of properties similar to minimal surfaces, and can be thought of as natural generalizations of the latter \cite{mots2}. Moreover, as Schoen and Yau discovered in their celebrated proof of the Positive Mass theorem \cite{SY1}, they have the striking feature that they arise as blowup sets of the Jang equation \cite{SY2}. Therefore, analysis of this blowup can lead to existence proofs for MOTS \cite{AEM,metzger}. 

Proving the existence of a MTS directly is more involved, because it entails control on the sign of certain quantities, which in general is a difficult problem \footnote{I thank Marc Mars and Michael Eichmair for their comments on this point.}. But given some physically motivated inputs, their appearance is natural. In Section \ref{more}, we briefly discuss some general conditions for the existence of MTS in initial data sets. Moreover, MTS appear in many spacetime contexts, such as cross-sections of event horizons in many exact solutions of the Einstein field equations, or even of more general horizons, like dynamical or trapping horizons. Indeed, the latter require conditions on the ingoing family of normal null geodesics (see, e.g., Ref. \cite{AK} for detailed definitions and properties). Indeed, in nearly all physically natural, concrete contexts in which such black hole horizon cross-sections appear, the underlying MOTS are also MTS.  

The rest of the paper is organized as follows. We first give some preliminary definitions in Section \ref{preliminaries} to set the conventions, general assumptions and notation, and a general proposition is proven which will be instrumental in the proof of our main results. The proof of the main result, Theorem \ref{mainsingularity2}, together with some simple consequences, are deferred to Section \ref{mainstuff}. In the final Section \ref{more}, we discuss additional conditions on initial data sets which imply the existence of MTS thereon.

\section{Preliminaries: Basic Definitions \& Results}\label{preliminaries}

 In what follows, we fix a {\em spacetime}, i.e, an $n$-dimensional, second-countable, connected, Hausdorff, smooth (i.e., $C^{\infty}$) time-oriented Lorentz (signature $(-,+, \ldots,+)$) manifold $M$ endowed with a smooth metric tensor $g$ and with $n \geq 3$. We assume that the reader is familiar with the basic definitions and results of global Lorentzian geometry and causal theory of spacetimes as found in the core references \cite{HE,wald,oneill,BE}, and in particular with the standard singularity theorems. We denote by $d: M \times M \rightarrow [0,+\infty]$ the (lower semicontinuous) Lorentzian distance function on $(M,g)$, and by $L_g(\gamma)$ the Lorentzian arc-length of a causal \footnote{The adjectives `causal' and `nonspacelike' as applied to curves (or curve segments) are used interchangeably.} curve segment $\gamma:[a,b] \rightarrow M$. All submanifolds of $M$ are embedded unless otherwise specified, and their topology is the induced topology. Finally, we follow the convention that causal vectors are always nonzero.
 
 We start by recalling a few standard definitions and results (cf. chapter 8 of Ref. \cite{BE} for examples and further discussion). Our intention here is merely to establish additional notation and terminology which will be used later on.  

\begin{definition}
\label{raysnlines}
A {\em future-directed timelike [resp. null] geodesic ray} in $(M,g)$ is a future-directed, future-inextendible timelike [resp. null] geodesic $\gamma : [0,b) \rightarrow M$ ($b\leq +\infty $) for which $d(\gamma(s),\gamma(t)) = L_g(\gamma |_{[s,t]})$ for all $s,t \in [0,b)$ with $s \leq t$.  In either case (i.e., timelike or null), $\gamma$ is also called a {\em future-directed nonspacelike geodesic ray}. 
\end{definition}

A {\em past-directed} timelike [resp. null, nonspacelike] geodesic ray or can be analogously defined. In what follows, unless explicitly stated otherwise, we shall always consider {\em future-directed} causal curves, and we henceforth drop explicit references to the causal orientation. According to the above definition, nonspacelike geodesic ray is characterized by the fact that it {\em maximizes the Lorentzian arc-length between any two of its points}. If $\gamma$ is a {\em null} ray, the condition that $d(\gamma(s),\gamma(t)) = L_g(\gamma |_{[s,t]})$ for all $s,t$ in the domain of $\gamma$ with $s \leq t$ is actually equivalent to the requirement that the image of $\gamma$ be {\em achronal} in $(M,g)$, i.e., no two of its points can be connected by a timelike curve segment.

Recall that an achronal set $A \subseteq M$ is {\em future-trapped} if $E^{+}(A) = J^{+}(A)\setminus I^{+}(A)$ is compact [a {\em past-trapped} set is defined time-dually]. 

Let $\S \subset M$ be a smooth, connected, spacelike, partial Cauchy hypersurface \footnote {Recall that a {\em partial Cauchy hypersurface} is by definition an acausal edgeless subset of a spacetime, which means in particular that it is a closed topological (i.e. $C^0$) hypersurface \cite{oneill}. In this paper, however, we always deal with {\em smooth} hypersurfaces.} (submanifold of codimension one), and a smooth, with each connected component of $\Sigma$ a compact (without boundary), spacelike submanifold of codimension two (loosely called {\em surface} in what follows) $\Sigma \subset \S$. We assume that $\Sigma$ is two-sided in $\S$. This means, in particular, that there are unique unit spacelike vector fields $N_{\pm}$ on $\Sigma$ normal to $\Sigma$ in $\S$.    

For simplicity we shall assume throughout this paper that $\Sigma$ is connected and {\em separates} $\S$, i.e., that $\S \setminus \Sigma$ is not connected. (This assumption is not essential for the results in this paper, however, for one can always focus on one connected component and consider a covering of $M$ in which this holds for that connected component of $\Sigma$. Now, in dealing with geodesic incompleteness one may as well work in covering manifolds.) Thus, $\S \setminus \Sigma$ is a disjoint union $\S_{+}\dot{\cup} \S_{-}$ of open submanifolds of $\S$ having $\Sigma$ as a common boundary. We shall loosely call $\S_{+}$ [resp. $\S_{-}$] the {\em outside} [resp. {\em inside}] of $\Sigma$ in $\S$. (In most interesting examples there is a natural choice for these.) The normal vector fields $N_{\pm}$ on $\Sigma$ are then chosen so that $N_{+}$ [resp. $N_{-}$] is outward-pointing [resp. inward-pointing], i.e., points into $\S_{+}$ [resp. $\S_{-}$].     

Let $U$ be the unique timelike, future-directed, unit normal vector field on $\S$. Then $K_{\pm} := U|_{\Sigma}+N_{\pm}$ are future-directed null vector fields on $\Sigma$ normal to $\Sigma$ in $M$. The {\em expansion scalars} of $\Sigma$ in $M$ are the smooth functions $\theta_{\pm}: \Sigma \rightarrow \mathbb{R}$ given by 
\begin{equation}
\label{expansion}
\theta_{\pm}(p) = - \langle H_p, K_{\pm}(p)\rangle _{p},
\end{equation}
for each $p \in \Sigma$, where $H_p$ denotes the mean curvature vector of $\Sigma$ in $M$ at $p$ \cite{oneill}, and we denote $g$ as $\langle \, , \, \rangle$ here and hereafter, if there is no risk of confusion. 

Henceforth, whenever we refer to a {\em surface $\Sigma$ contained in a partial Cauchy hypersurface $\S$}, all the above conventions will be understood. 

Physically, the expansion scalars measure the divergence of light rays emanating from $\Sigma$. If $\Sigma$ is a round sphere in a Euclidean slice of Minkowski spacetime, with the obvious choices of inside and outside, we have $\theta_{+}>0$ and $\theta_{-} < 0$. One also expects this to be the case if $\Sigma$ is a ``large"  sphere in an asymptotically flat spacetime. But in a region of strong gravity one expects instead that we have both $\theta_{\pm} < 0$, in which case $\Sigma$ is a {\em closed (future) trapped surface}. $\Sigma$ is a {\em marginally outer trapped surface} (MOTS) is $\theta_{+} = 0$. 

We also adopt the following slightly less standard definition (conf. \cite{wald}, p. 310): 
\begin{definition}
\label{MTS}
 A (future) {\em marginally trapped surface} (MTS) is a smooth, closed, codimension two spacelike manifold $\Sigma \subset M$ with trivial normal bundle in $M$ for which $\theta_{\pm} \leq 0$.   
\end{definition}

Note that a MTS may be a MOTS, but there is an additional requirement on the inner expansion scalar $\theta_{-}$ {\em There are slightly different definitions of MTS in the literature. The condition given here for $\theta_{-}$ is weaker than that imposed for cross-sections of dynamical and trapping horizons, which require a strict inequality $\theta_{-} <0$ (conf. \cite{AK}).}. It is well known (see, e.g, p. 310 of \cite{wald}) that either MTS or MOTS remain inside the black hole region, provided they are in a strongly asymptotically predictable spacetime in which the null convergence condition holds. 

{\em In this paper, unless otherwise specified, we shall only deal with MTS}. Indeed, we shall consider the following additional notion: 
 
\begin{definition}
\label{genericMTS}
 A MTS $\Sigma$ is {\em generic} (in $(M,g)$) if any future-directed, future-inextendible, or past-directed, past-inextendible null geodesic $\eta:[0,a) \rightarrow M$ ($0<a\leq +\infty$) starting at $\Sigma$ and normal to $\Sigma$ at $\eta(0)$ satisfies the generic condition, i.e., at some point $p$, and for some vector $v$ normal to $\eta'(p)$, $\langle v,R(v,\eta')\eta' \rangle \neq 0$.      
\end{definition}

The generic condition formulated above is a mild constraint which ensures (see, e.g., Prop. 2.11 in \cite{BE}) that there is non-zero tidal acceleration somewhere along each null geodesic $\eta$ normal to $\Sigma$. This will occur, for example (see Prop. 2.12 in \cite{BE}), if $Ric(\eta',\eta') \neq 0$ somewhere along $\eta$, which in turn, if the Einstein field equations hold for $(M,g)$, will likely happen whenever $\eta$ crosses (or is part of) some matter-energy cluster of positive density.  

A null geodesic $\eta:[0,a) \rightarrow M$ normal to a MTS $\Sigma$ is said to be {\em outward-pointing} [resp. {\em inward-pointing}] if $\eta'(0)$ is parallel to $K_{+}(\eta(0))$ [resp. $K_{-}(\eta(0))$]. 

The proofs of our main results will be based on the following

\begin{proposition}
\label{mainproposition}
Let $\Sigma \subset M$ be a surface contained in a partial Cauchy hypersurface $\S$. Then, at least one of the following alternatives occurs: 
\begin{itemize}
\item[i)] $\Sigma$ is a future-trapped set, 
\item[ii)] there exists an inward-pointing, future-directed geodesic ray starting at $\Sigma$ and normal to $\Sigma$, 
\item[iii)] there exists an outward-pointing, future-directed geodesic ray starting at $\Sigma$ and normal to $\Sigma$. 
\end{itemize}
\end{proposition} 

{\em Proof.} Fix a complete Riemannian metric $h$ on $M$ with distance function $d_h$.

If $E^{+} (\Sigma) = \Sigma$, we are done, since $\Sigma$ is compact. Thus, suppose $E^{+} (\Sigma) \neq \Sigma$. Let $(q_n)$ be any sequence in $E^{+} (\Sigma)\setminus \Sigma$. For each $n \in \mathbb{N}$, there exists a future-directed, future-inextendible causal curve $\gamma_n:[0, +\infty) \rightarrow M$ parametrized by $h$-arc length, such that $\gamma_n(0) \in \Sigma$ and $\gamma_n(t_n) = q_n$, and $\gamma_n(0,t_n] \subset E^{+} (\Sigma)\setminus \Sigma$ for some $t_n \in (0, +\infty)$. 

Since $\Sigma$ is compact, passing to a subsequence if necessary we can assume that $\gamma_n(0) \rightarrow p \in \Sigma$. By the Limit Curve Lemma, we can assume that there exists a future-directed, future-inextendible causal curve $\gamma:[0,+\infty) \rightarrow M$ with $\gamma(0) =p$ and such that $\gamma_n|_{C} \rightarrow \gamma|_{C}$ $h$-uniformly in compact subsets. 

Assume that $E^{+} (\Sigma)$ is not compact. Then, by the Hopf-Rinow theorem in $(M,h)$, it is either not closed, or not bounded in $d_h$. Suppose first that it is not bounded. In that case, the sequence $(q_n)$ could be chosen as diverging to infinity \footnote{We say that a sequence $(p_n)_{n \in \mathbb{N}}$ in $M$ {\em diverges to infinity} if given any compact subset $C \subseteq M$, only finitely many elements of the sequence are contained in $C$ (conf. \cite{BE}, ch. 8).}; then, the sequence $(t_n)$ is not bounded above, and we can assume that $t_n \rightarrow +\infty$. 

Fix a number $t >0$. Eventually, $t_n > t$, in which case $\gamma_n(t) \in J^{+}(\Sigma)$, and hence $\gamma_n(t) \in E^{+} (\Sigma)$, for otherwise $\gamma_n (t) \in I^{+}(\Sigma)$ and therefore $q_n \in I^{+}(\Sigma)$, a contradiction. Similarly, we cannot have $\gamma(t) \in I^{+}(\Sigma)$; therefore $\gamma(t) \in E^{+} (\Sigma)$, and since    $\gamma(t) \notin \Sigma$ from the acausality of $\S$, we conclude that $\gamma \subset E^{+} (\Sigma)$. In particular, there cannot exist any timelike curve from $\Sigma$ to any point of $\gamma$, so the latter curve must be a reparametrization of a future-directed, null geodesic ray normal to $\Sigma$ as desired (cf. Theorem 51, p. 298 of Ref. \cite{oneill}). Since $\Sigma$ is two-sided, this ray will be either inward-pointing or outward-pointing.

Now suppose that $E^{+} (\Sigma)$ is not closed. Then we can take $(q_n)$ such that $q_n \rightarrow q$, with $q \notin E^{+} (\Sigma)$. If $(t_n)$ is not bounded above, then the argument proceeds as in the previous situation. Thus, assume that $(t_n)$ is bounded above. In that case, up to passing to a subsequence we can assume that $t_n \nearrow t_0 \in [0,+\infty)$, hence $q = \gamma(t_0) \in  J^{+}(\Sigma)$. If $q \in I^{+}(\Sigma)$, which is open, then eventually so does $q_n$, which is absurd. Thus, $q \in E^{+} (\Sigma)$, again a contradiction. Therefore, $(t_n)$ is not bounded above. 

The remaining possibility is that $E^{+} (\Sigma)$ is compact, in which case $\Sigma$ is future-trapped. 

\qcd

\begin{remark}
{\em Clearly, a time-dual statement of Proposition \ref{mainproposition} also holds.}
\end{remark} 

\begin{remark}
\label{sigmaray}
{\em Note that the rays constructed in the case $(ii)$ or $(iii)$ in Proposition \ref{mainproposition} are actually more than just normal rays, but each of them is actually a {\em $\Sigma$-ray} (conf. Definition 14.4 in Ref. \cite{BE}), i.e. the length of any segment starting at $\Sigma$ up to any point along the ray realizes the Lorentzian distance from $\Sigma$ to that point. In particular, there exists no timelike curve from $\Sigma$ to the ray and no focal points to $\Sigma$ along the ray.}
\end{remark}

\section{Main Results \& Consequences}\label{mainstuff}


The following technical Lemma, proven by Guimaraes \cite{guima} for solutions of Riccati-type ODE's will be useful:

\begin{lemma}
\label{guimalemma}
Let $u:[0, + \infty) \rightarrow \mathbb{R}$ be a continuous function. Then, for any $c>0$,
\[
\limsup_{s \rightarrow + \infty} \left( u(s) + c \, \int^{s}_0 \left(u(t)\right)^2 dt \right) \geq 0,
\]
with equality iff $u \equiv 0$. 
\end{lemma}

We are now ready to state and prove our main theorem. 
 
\begin{theorem}
\label{mainsingularity2}
Let $(M,g)$ be a spacetime of dimension $n \geq 3$ satisfying the the following requirements:
\begin{itemize}
\item[(1)] $(M,g)$ contains no closed timelike curve.
\item[(2)] $(M,g)$ obeys the timelike convergence condition, i.e. its Ricci tensor satisfies $Ric(v,v)\geq 0$, for all timelike vectors $v$.
\item[(3)] $(M,g)$ contains a generic MTS $\Sigma$ contained in a partial Cauchy hypersurface.
\item[(4)] The nonspacelike generic condition holds in $(M,g)$.
\end{itemize}
Then $(M,g)$ is nonspacelike geodesically incomplete
\end{theorem}

{\em Proof.} Suppose first that $\Sigma$ is not a future-trapped set. Then, by Prop. \ref{mainproposition} there exists a future-directed, affinely parametrized null geodesic $\eta:[0, a) \rightarrow M$ normal to $\Sigma$. Because we are assuming that $\Sigma$ is a MTS and not only a MOTS, the argument is similar either for an outward- or inward-pointing geodesic, so we may take it to be outward-pointing for definiteness. Suppose $\eta$ is future-complete, so that we put $a= +\infty$. Since $\eta$ is a $\Sigma$-ray (conf. Remark \ref{sigmaray}) a standard analysis (see Ch. 12 of \cite{BE}) there exists a globally defined Lagrange tensor field $A(t)$ on $\eta$ such that $b(t) = A'(t)A^{-1}(t)$ obeys the Riccati-type equation
\begin{equation}
\label{riccati}
b' + b^2 + {\cal R} =0,
\end{equation}
where ${\cal R}(t) : N(\eta(t)) \rightarrow N(\eta(t))$ takes the values of $R(v,\eta')\eta'$ on normal vectors $v \bot \eta(t)$ (taken modulo $\eta'$), $N(\eta(t))$ being their span, and so that $\theta_b(t) \equiv tr(b(t))$ satisfies $\theta_b(0) = \theta_{+}(\eta(0)) \leq 0$. Now, tracing Eq. (\ref{riccati}) we obtain the Raychaudhuri equation
\begin{equation}
\label{raychau}
\theta'_b + \frac{1}{n-2} \theta^2_b = - Ric(\eta',\eta') - \sigma ^2,
\end{equation}    
where the shear scalar $\sigma$ is the trace of the square of the trace-free part of $b$. If we define $f:[0,+\infty) \rightarrow \mathbb{R}$ by
\begin{equation}
\label{convenience}
f(t) = - Ric(\eta'(t),\eta'(t)) - \sigma ^2(t),
\end{equation}
the assumption on the Ricci tensor implies that $f\leq 0$, so that integrating Eq. (\ref{raychau}) from $0$ up to an arbitrary $s$, we get
\[
\limsup_{s \rightarrow + \infty} \left(\theta_b(s) + \frac{1}{n-2}\int^{s}_0 \left(\theta_b(t)\right)^2 dt \right) = \theta_b(0) + \limsup_{s \rightarrow + \infty} \int^s_0 (f)dt \geq 0,
\]
and using Lemma \ref{guimalemma} (for $c= 1/(n-2)$) we conclude that $\theta_b \equiv 0$, and again by Eq. (\ref{raychau}), that $f \equiv 0$ (and also that we must have $\theta_{+}(\eta(0)) =0$!). But then $b \equiv 0$ along $\eta$ and hence ${\cal R} \equiv 0$ by Eq. (\ref{riccati}), contradicting the generic condition on normal geodesics. 

Therefore, either $\eta$ is future incomplete and we are done, or $\Sigma$ is a future-trapped set. In the latter case, the Hawking-Penrose singularity theorem now yields the conclusion, so the proof is complete. 
\qcd

In Section 3 of Ref. \cite{EGP}, the authors call an {\em initial data singularity theorem} any result which proves the existence of closed trapped surfaces in initial data sets from suitable conditions on the geometry and/or energy-momentum density. Of course, what is really relevant for singularity theorems are trapped {\em sets}, closed trapped surfaces being just one class of such sets when some additional conditions hold (strong causality plus the null convergence condition suffice in this case). However, they are particularly important for, say, numerical Relativity, because the signs of the expansion scalars can be ascertained in a compact region of the initial data underlying manifold, and can be expressed solely in terms of the initial data. MTSs or MOTS also are likewise quasi-locally detectable. In particular, the proof of Theorem \ref{mainsingularity2} actually means that any proof of the existence of a MTS in initial data sets can also be viewed as an initial data singularity theorem in this sense:   

\begin{theorem}
\label{mainsingularity3}
Let $(M,g)$ be a spacetime of dimension $n \geq 3$ in which the null convergence condition holds, and which contains a generic MTS $\Sigma$ in a partial Cauchy hypersurface. Then, either $(M,g)$ is null geodesically incomplete, or $\Sigma$ is a future-trapped set. 
\end{theorem}

Theorem \ref{mainsingularity3} and the standard Penrose theorem \cite{P}, the analogue for MTS of the singularity theorem in \cite{EGP} then follows:

\begin{corollary}
\label{mainsingularity4}
Let $(M,g)$ be a globally hyperbolic spacetime of dimension $n \geq 3$ in which the null convergence condition holds, and which contains a generic MTS $\Sigma$ in a noncompact Cauchy hypersurface $\S$. Then $(M,g)$ is null geodesically incomplete.  
\end{corollary}

{\em Proof.} Suppose not. Then, by Theorem \ref{mainsingularity3}, $\Sigma$ is future-trapped. A standard argument (see e.g. , Theorem 61, pg. 437, of \cite{oneill}) establishes that $\S$ and $E^{+}(\Sigma)$ are homeomorphic, a contradiction. 

\qcd

\begin{remark}
{\em The results in Theorem \ref{mainsingularity2} remains valid if we have a generic MOTS instead of a generic MTS, provided that the MOTS {\em bounds}, i.e., provided that $\overline{\S_{-}}$ is compact. Note that $(1),(2)$ and $(4)$ together imply that $(M,g)$ is strongly causal, and in this case, it is possible to show that the existence of an inward-pointing $\Sigma$-ray is ruled out, so by Proposition \ref{mainproposition}, either $\Sigma$ is trapped or else the ray must be outward-pointing, and in this case the ray will be incomplete. Details will appear elsewhere. Similarly, Theorem \ref{mainsingularity3} is still valid for a generic bounding MOTS provided we add the hypothesis that $(M,g)$ is strongly causal.}
\end{remark}
\begin{remark}
{\em Regarding the previous Remark, a celebrated result by Schoen and Yau establish that compact MOTS appear naturally for large concentrations of matter \cite{SY3}. In the case when $\S$ is simply-connected, such MOTS will bound \footnote{I thank Greg Galloway for suggesting this example, together with Ref. \cite{SY3}.}.}
\end{remark} 
\section{Some remarks on the existence of a MTS in initial data sets}\label{more}

An important part of the interest in MOTS as quasi-local versions of black hole horizons springs from the fact that their existence and properties can be studied at the initial data set level. Our goal in this Section is to make a few general remarks on the existence of MTS in such initial data sets. 

Fix, for the remaining of this Section an {\em ($n$-dimensional) initial data set}, by which we mean a triple $(N,h,K)$, where $(N,h)$ is a smooth $n$-dimensional Riemannian manifold and $K$ is a smooth, symmetric $(0,2)$ tensor field over $N$. We can then define a real-valued function $\rho$ and a 1-form $J$ on $N$ by 
\begin{eqnarray}
\label{constraint1}
\rho &:=& \frac{1}{2} \left( R_N - |K|^2 + \left( \mbox{tr} _N K \right) ^2 \right); \\
\label{constraint2} J &:=& {\mbox div} _N \left( K - \left( \mbox{tr} _N K \right) h \right),
\end{eqnarray}  
where $R_N$ denotes the scalar curvature of $N$. 

Of course, the definitions above are purely geometric, but they acquire physical meaning when applied in the initial-value formulation of General Relativity. In this setting, $N$ is to be thought of as an embedded spacelike hypersurface in an $(n+1)$-dimensional spacetime $(M,g)$, with $h$ being the induced metric and $K$ being the second fundamental form, taken with respect to the unique future-directed, unit timelike normal vector field $U$ over $N \hookrightarrow M$. Moreover, $(M,g)$ is assumed to satisfy the Einstein field equation 
\begin{equation}
\label{EFE}
Ric_M - \frac{1}{2} R_M g + \Lambda g = T,
\end{equation}
for a suitable $(0,2)$ symmetric energy-momentum tensor $T$, and a (possibly vanishing) cosmological constant $\Lambda$. One can check that the Gauss-Codazzi equations for the embedding $N \hookrightarrow M$ imply that the initial data set $(N,h,K)$ automatically satisfies Eqs. (\ref{constraint1}) and (\ref{constraint2}) with the identifications 
\begin{eqnarray}
\rho &\equiv & T(U,U) + \Lambda, \\
J &\equiv & -T(U, \cdot ).
\end{eqnarray} 
In this context, one may consider how to express the expansion scalars $\theta_{\pm}$ of a surface $\Sigma^{n-1} \subset N$, viewed as a codimension two submanifold of $M$, in terms of initial data quantities. With the conventions of the previous Section, a straightforward computation gives 
\begin{equation}
\label{thetainitial1}
\theta_{\pm} = \mbox{tr}_{\Sigma}K \pm H_{\Sigma},
\end{equation}
where $H_{\Sigma}$ denotes the mean curvature of $\Sigma$ as a (codimension one) submanifold of $N$ with respect to the outward-pointing normal $N_{+}$, and given any orthonormal frame $\{E_1, \ldots, E_{n-2} \}$ on $\Sigma$, 
\[
\mbox{tr}_{\Sigma}K \equiv \sum_{i=1}^{n-2} K(E_i,E_i).
\]
Using Eq. (\ref{thetainitial1}), we conclude that {\em a MOTS is a MTS iff it is mean-convex, i.e., iff $H_{\Sigma}\geq 0$}. In Ref. \cite{EGP}, it is discussed how a non-trivial topology of $N$ can give rise to the existence of MOTS, but to the best of this author's knowledge, there is no result in the literature ensuring the existence of a {\em mean-convex} MOTS solely under geometrical conditions of this sort.

But given some physically motivated {\em extrinsic} conditions on the initial data, one can say a little more. The condition that $\Sigma$ be a MTS, $\theta_{\pm} \leq 0$, implies, from Eq. (\ref{thetainitial1}), that  
\begin{equation}
\label{thetainitial2}
\mbox{tr}_{\Sigma}K \leq H_{\Sigma} \leq -\mbox{tr}_{\Sigma}K,
\end{equation}
so in particular we must have $\mbox{tr}_{\Sigma}K \leq 0$. Now, adapting a definition given by Galloway in Ref. \cite{galloway2}, we can introduce the following notion.
\begin{definition}
\label{nonexpanding}
$N$ is {\em non-expanding in all directions} at a point $p \in N$ if $K$ is negative semi-definite on the tangent space $T_pN$. A set $A \subseteq N$ is {\em non-expanding} if $N$ is non-expanding in all directions at every $p \in A$. 
\end{definition}
The intuitive meaning of this definition becomes clear if $N$ describes (a portion) of the Universe at a given ``instant'': one would expect that $N$ is non-expanding in all directions at all points of a region of $A \subseteq N$ undergoing gravitational collapse. In fact, with our sign conventions a constant $t$ surface in a Robertson-Walker spacetime $(I \times \S, -dt^2 \oplus f^2(t)d\sigma^2)$ is non-expanding iff $f'(t)\leq 0$ for that surface.  However, for {\em time-symmetric} initial data, i.e., when $K \equiv 0$, $N$ will be non-expanding in all directions at {\em all} points. Then, $N$ would be non-expanding even if there is no gravitational collapse. Moreover, since this is an extrinsic condition on the initial data, even in spacetimes describing gravitational collapse this condition might not hold for any initial data set thereof.    

Therefore, as long as the additional condition of non-expansion holds for appropriate subsets in the initial data underlying manifold, one can benefit from extant results on the existence of MOTS and/or minimal surfaces in initial data sets (see Section 4 of \cite{EGP} and references therein for some of these). In fact, we have:
\begin{proposition}
\label{simple} 
Suppose that $\Sigma$ is either a MOTS or a minimal surface in $(N,h)$, contained in a non-expanding set $A \subseteq N$. Then $\Sigma$ is a MTS. 
\end{proposition}
{\em Proof.} If $\Sigma$ is a minimal surface, this follows immediately from the definition, and from Eqs. (\ref{thetainitial1}) with $H_{\Sigma} \equiv 0$. If $\Sigma$ is a MOTS, then Eqs. (\ref{thetainitial1}) imply that $H_{\Sigma} = -\mbox{tr}_{\Sigma}K$, and therefore 
\[
\theta_{-} = 2\mbox{tr}_{\Sigma}K \leq 0.
\]

\qcd

\begin{acknowledgements}
I wish to thank Greg Galloway for his careful reading of the first version of this paper, and for his valuable comments thereon. 
\end{acknowledgements}


%
\end{document}